\documentclass[twocolumn,apj]{emulateapj}

\slugcomment{}
\usepackage{color}

\shorttitle{Balmer continuum in flares}
\shortauthors{Heinzel and Kleint}

\begin{document}


\title{Hydrogen Balmer Continuum in Solar Flares Detected by the\\
    Interface Region Imaging Spectrograph (IRIS)}


\author{P. Heinzel}
\affil{Astronomical Institute, Academy of Sciences of the Czech Republic\\
    Fri\v{c}ova 298, 25165 Ond\v{r}ejov, Czech Republic}

\author{L. Kleint}
\affil{University of Applied Sciences and Arts Northwestern Switzerland\\
Bahnhofstrasse 6, 5210 Windisch, Switzerland}



\begin{abstract}
We present a novel observation of the white-light flare (WLF) continuum, which was significantly enhanced during
the X1 flare on March 29, 2014 (SOL2014-03-29T17:48). Data from the Interface Region Imaging Spectrograph (IRIS) in its NUV channel show that at the peak of the continuum enhancement, the contrast at the quasi-continuum window above 2813 \AA\ reached 100 - 200 \%
and can be even larger closer to the \ion{Mg}{2} lines. This is fully consistent with the hydrogen recombination
Balmer continuum emission, which follows an impulsive thermal and non-thermal ionization caused by
the precipitation of electron beams through the chromosphere. However, a less probable photospheric
continuum enhancement cannot be excluded. The light curves of the Balmer continuum have an impulsive
character with a gradual fading, similar to those detected recently in the optical region on Hinode/SOT.
This observation represents a first Balmer-continuum detection from space far beyond the Balmer
limit (3646 \AA), eliminating seeing effects known to complicate the WLF detection.
Moreover, we use a spectral window so far unexplored for flare studies, which provides the potential to study the Balmer continuum, as well as many metallic lines appearing in
emission during flares. Combined with future ground-based observations of the continuum near
the Balmer limit, we will be able to disentangle between various scenarios of the WLF origin. 
IRIS observations also provide a critical quantitative measure of the energy radiated in the Balmer
continuum, which constrains various models of the energy transport and deposition during flares. 
\end{abstract}





\section{Introduction}

Radiation emitted during a solar flare from the lower solar atmosphere (chromosphere and photosphere)
represents a significant, if not dominant, portion of the total energy deposited in those
layers by various mechanisms. Considering the standard collisional thick-target model \citep[CTTM, ][]{1971SoPh...18..489B}, 
the electron beam transports the energy from the coronal reconnection site down to the lower atmosphere,
where its energy is dissipated. The heated atmosphere produces strongly enhanced radiation in many
spectral lines and various continua, but the emission also comes from direct non-thermal collisional excitations
and ionizations of the plasma by the beam. The most spectacular visible-continuum emission is known as the
white light flare (WLF), first detected by \citet{1859MNRAS..20...13C}. Two mechanisms are currently considered
to be responsible for this optical emission \citep[see, e.g.,][]{2007ASPC..368..417D}: (i) photospheric continuum enhancement as a signature of the 
temperature increase below the temperature minimum region (mainly the H$^-$ continuum), and (ii)
hydrogen recombination continua (Paschen, Balmer) produced in the chromosphere.
The radiation in the latter continua emitted upwards can be directly detected, while the
downward component is supposed to heat the photosphere,
the so-called 'back-warming' \citep{1989SoPh..124..303M}. In EUV, the hydrogen emits a strong 
Lyman continuum below 912 \AA\ and
this was recently well detected by the SDO/EVE spectrometer \citep{2012ApJ...748L..14M,2014arXiv1406.7657M}.  A new
attempt to estimate the energy contained in the optical continua was recently made using Hinode/SOT \citep{2013ApJ...776..123W,2014ApJ...783...98K,2014arXiv1406.7657M}. The spectral coverage is rather poor, but the data allow 
to estimate the total power by fitting the continuum with the black-body curve. However, such a black-body fit leads to a low enhancement in the
Balmer continuum, contrary to numerical simulations of the hydrogen recombination continua, which predict much higher emission. Therefore,
observations of the Balmer continuum during flares are highly desirable in order to
set-up constraints on various mechanisms producing the WLFs.
\begin{figure*}[htb]    \centering 
   \includegraphics[width=.85\textwidth, bb=30 50 500 500]{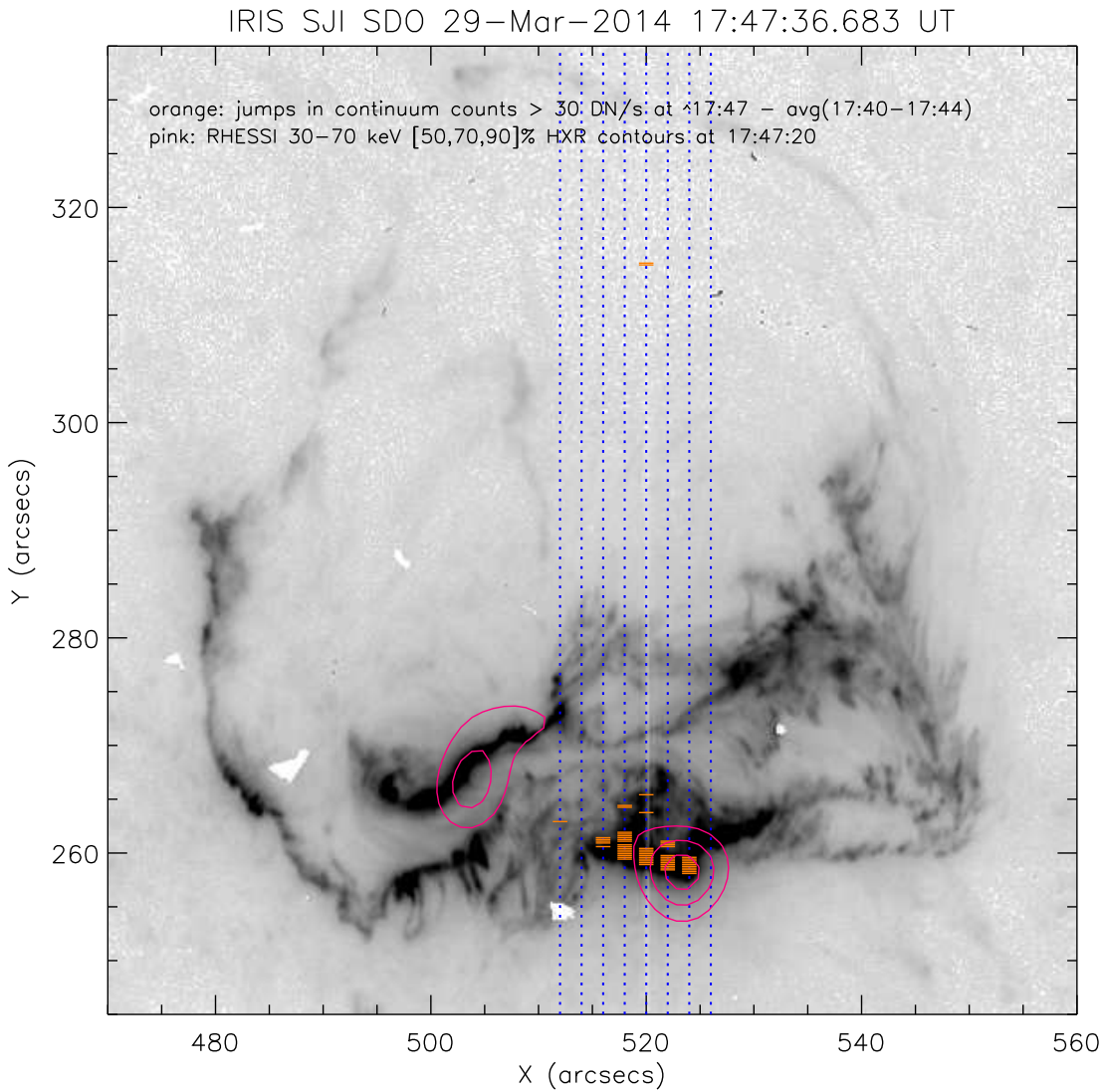}
   \caption{IRIS 1400 SJI with intensity reversed showing the flare ribbons in black. The intersections of the horizontal orange lines with the vertical IRIS slit positions (dotted blue) mark positions where the continuum increased significantly. Pink RHESSI 30--70 keV HXR contours are drawn for reference.}
         \label{fig1}
  \end{figure*}

Only few detections of the Balmer-continuum brightening during flares are described in the literature. 
All have been made from the ground and only
close to the Balmer limit at 3646 \AA. In some cases, the Balmer jump was detected, in others a smooth transition
from the so-called 'blue continuum'  \citep{1985A&A...152..165D} to the Balmer continuum was observed \citep[see summary by][]{1989SoPh..121..261N,2014ApJ...783...98K}.
Quite recently, this was also detected by \citet{kotrc2014} who measured the flux from the whole
flare area. On the other hand, \citet{1976GAM.....8.....S} claims in his book that no Balmer
continuum was detected in many flare spectra collected at the Ond\v{r}ejov observatory. The detectability
limit may be related to the photographic technique used, but others detected non-negligible enhancements \citep[see e.g.,][]{1981ApJ...248L..45Z,1982SoPh...80..113H}.

In this Letter we present a novel observation of the Balmer continuum, obtained from space
and far beyond the Balmer limit. We therefore eliminate the seeing effects known to complicate WLF
observations and simultaneously consider an unexplored spectral window. This was achieved using 
recent Interface Region Imaging Spectrograph  \citep[IRIS, ][]{2014SoPh..289.2733D} observations of a large X-class flare, and namely in the near-UV (NUV) channel suitable for the detection of the
Balmer continuum. In this wavelength range, the expected WLF contrast is much higher compared to the
visible part of the spectrum.
Moreover, the IRIS slit is continuously scanning the flare area and thus we get an uninterrupted time
evolution of the Balmer-continuum enhancement which we compare with RHESSI and GOES lightcurves.

\begin{figure*}[htb]    \centering 
   \includegraphics[width=.9\textwidth]{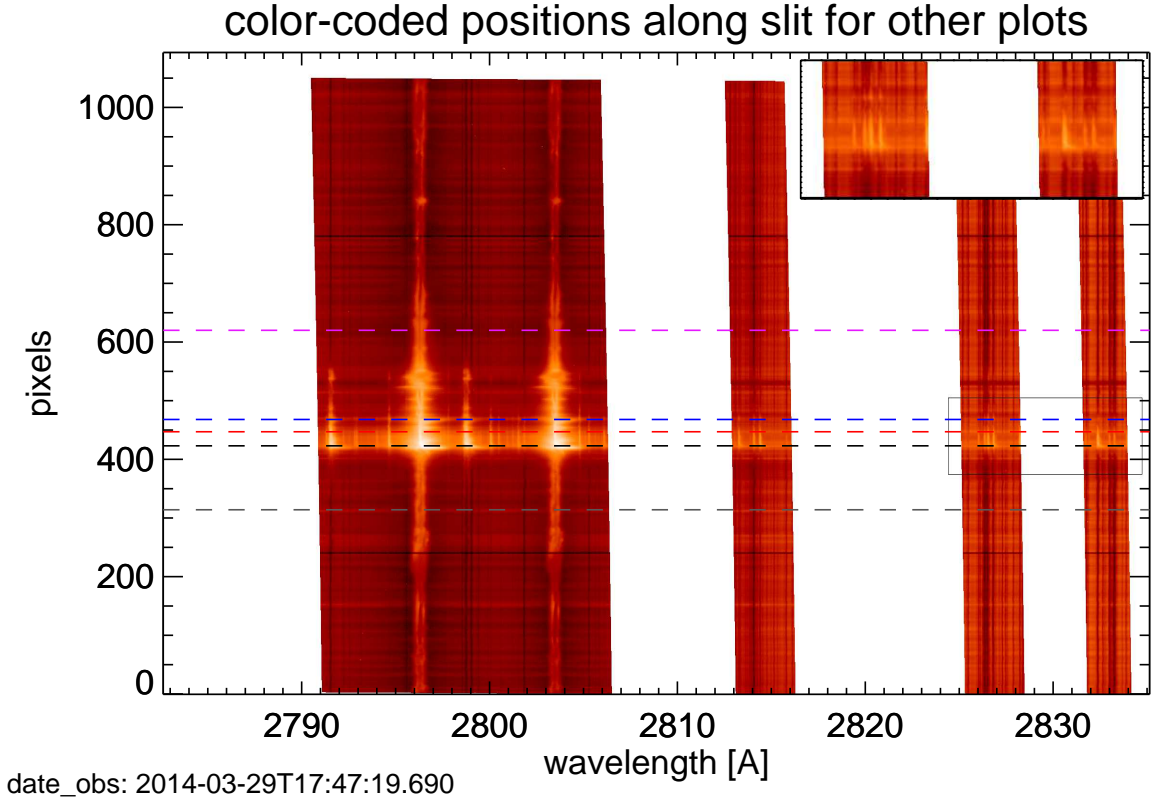}
   \caption{IRIS NUV spectrum showing the intensity enhancement throughout the whole spectrum, including the continuum, which was zoomed in on the top right. Reference positions for the following plots are indicated by horizontal lines.}
         \label{fig2}
  \end{figure*}

\section{IRIS Observations}
The well-observed X1 flare on March 29, 2014 (SOL2014-03-29T17:48) occurred in NOAA AR 12017, with GOES X-ray flux enhancements starting around 17:35 UT and peaking at 17:48 UT. For a complete timeline of the event, see \citet{kleint2014}. IRIS was carrying out a coordinated observing program with the Dunn Solar Telescope and Hinode from 14:09 -- 17:54 UT and captured the whole impulsive phase and part of the decay phase of the flare. The center of the flare was approximately located at coordinates (265\arcsec\ N, 515\arcsec\ W), which corresponds to (10.3$^\circ$ N, 32.9$^\circ$ W) and $\mu = \cos \theta$ = 0.83.

The IRIS observation consisted of an 8-step raster in steps of 2\arcsec. At each raster step, a far-UV spectrum (FUV), a near-UV spectrum (NUV), and a slitjaw image (SJI) was taken. The total field-of-view was 14\arcsec $\times$ 174\arcsec\ for the spectra and $\sim$174 $\times$174\arcsec \, for each SJI image (see Fig.~\ref{fig1}). 
The nominal exposure time of each image was 8 seconds, which, including overhead, yielded a cadence of 75 seconds per raster. During the flare, the NUV and SJI had automatic exposure control enabled, reducing their exposure times to minimize saturation of the images. The dispersion of the FUV and NUV spectra is 25.46 m\AA/px for this observation and the plate scale is 0.166\arcsec/px.

Here, we focus on the NUV spectra, which contain the two strong lines \ion{Mg}{2}  {\em h} and {\em k} and their extended wings with many blend lines. The NUV spectra may record wavelengths from 2783 -- 2835 \AA, but usually the detector is only read out partially to save time and downlink volume. For these observations, the ``Flare linelist'' was used, which saves four spectral ranges: $\sim$2791--2806 \AA, $\sim$2813--2816 \AA, $\sim$2825--2828 \AA, and $\sim$2831--2834 \AA, shown in Fig.~\ref{fig2} where white indicates parts of the detector that were not read out or saved. We used the calibrated Level 2 data, which include corrections for dark current, flatfield, geometry, but do not include an absolute radiometric calibration.

\section{Analysis of the IRIS Flare Spectra}

We analyzed the IRIS data for flare-related brightness increases in the continuum, defined as far wing of the \ion{Mg}{2} line where no spectral lines are visible. 
This is usually called a 'quasi-continuum' and is well visible in a broader spectrum taken by HRTS-9 where the quasi-continuum peaks between the \ion{Mg}{2} $h$ and strong \ion{Mg}{1} line \citep{2008ApJ...687..646M}.
Figure~\ref{fig1} shows a (negative) SJI 1400 image, with the flare ribbons appearing in black. The white particles are dust on the camera. The eight vertical dotted lines mark the IRIS slit positions of the raster. The intersections of small orange horizontal lines with the slits mark locations where the continuum counts increased by at least 30 DN/s at $\sim$17:47 UT compared to the average counts from $\sim$17:40 -- 17:44 UT. The exact times are different for each slit position because of the duration of the raster, e.g. for slit position 1 we looked at counts at 17:46:51 UT - avg(17:40:36 -- 17:44:21 UT), while for position 8 it was 17:47:56 UT - avg(17:41:42 -- 17:45:27 UT). The continuum enhancements clearly coincide with the flare ribbon.
The pink contours show the RHESSI 30--70 keV HXR emission at [50, 70, 90]\%, for context. 
While the location of HXR and continuum brightening agrees in general, a more detailed analysis that includes the motion of the HXR footpoints and the exact timing of each raster step will be carried out in the future to explore a possible relationship of the continuum brightness and the HXR emission.

Figure~\ref{fig2} shows the NUV spectrum at 17:47:19 UT with the two bright \ion{Mg}{2} {\em k} and {\em h} lines and three spectral regions further in the $h$-wing to the red. An enhancement of brightness is clearly visible around y=420 px, not only in the spectral lines, but also in the continuum. The inset on the top right shows a magnified part of the spectrum denoted by the box. The horizontal color-coded lines refer to positions along the slit, which are used for the following analysis.
\begin{figure*}[htb]    \centering 
   \includegraphics[width=\textwidth]{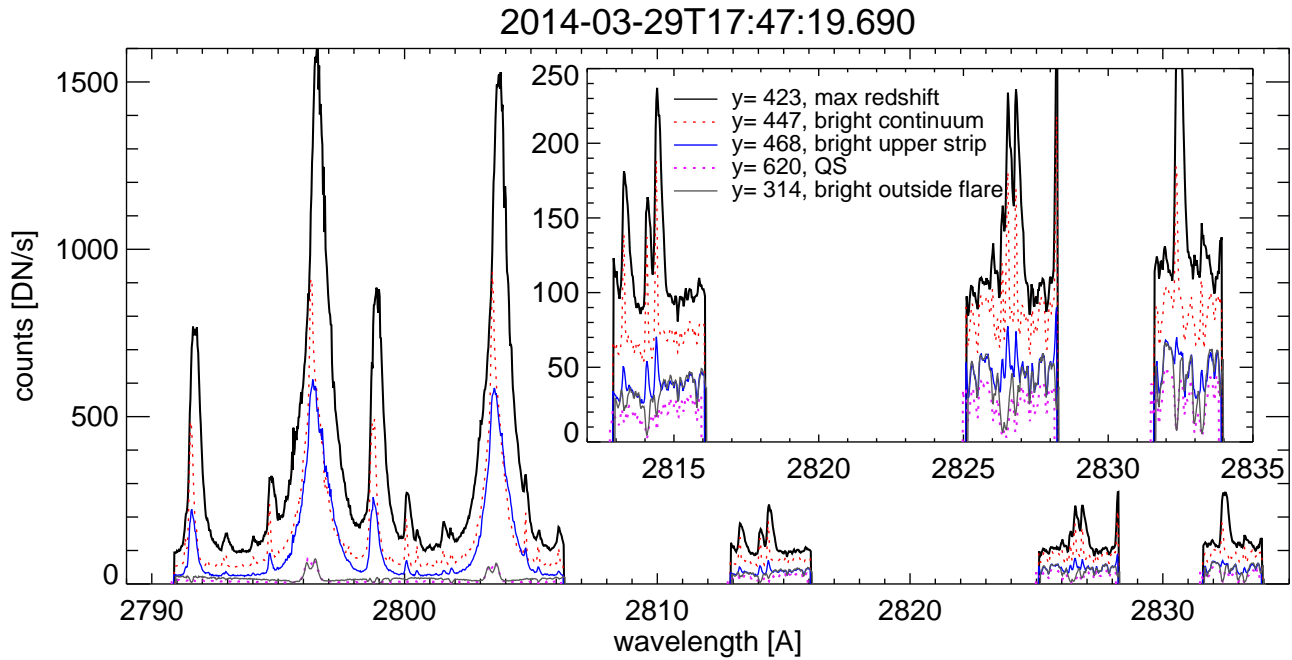}
   \caption{Example spectra during the flare at the five different positions along the slit indicated in Fig.~\ref{fig2}. The continuum values are higher in the black and red spectra, which are from the flare footpoint. Note that the QS quasi-continuum increases with increasing wavelength which is due to the photospheric wing of the  \ion{Mg}{2} $h$ line.}
         \label{fig3}
  \end{figure*}

Figure~\ref{fig3} shows the NUV spectrum at the selected locations from Fig.~\ref{fig2}. At the brightest positions along the slit (black line) the continuum counts are seen to increase by a factor of 2 - 3 compared to the quiet Sun (pink dotted line). The other spectra were chosen to show other bright flare regions (red dotted and blue lines), and a bright feature outside of the flare (grey line). The inset of the figure shows a magnification of the continuum regions. While many spectral lines go into emission during the flare (black, red, and blue spectra), the black and red spectra also show an increased continuum, compared to the reference spectra from the quiet Sun (pink), and the bright feature (grey). Comparing the blue and the grey spectra, it seems that the only difference are the lines in emission, while the continuum is similar. In other words, the continuum between lines is not affected by the line emission itself.

\begin{figure*}[htb]    \centering 
   \includegraphics[width=.9\textwidth]{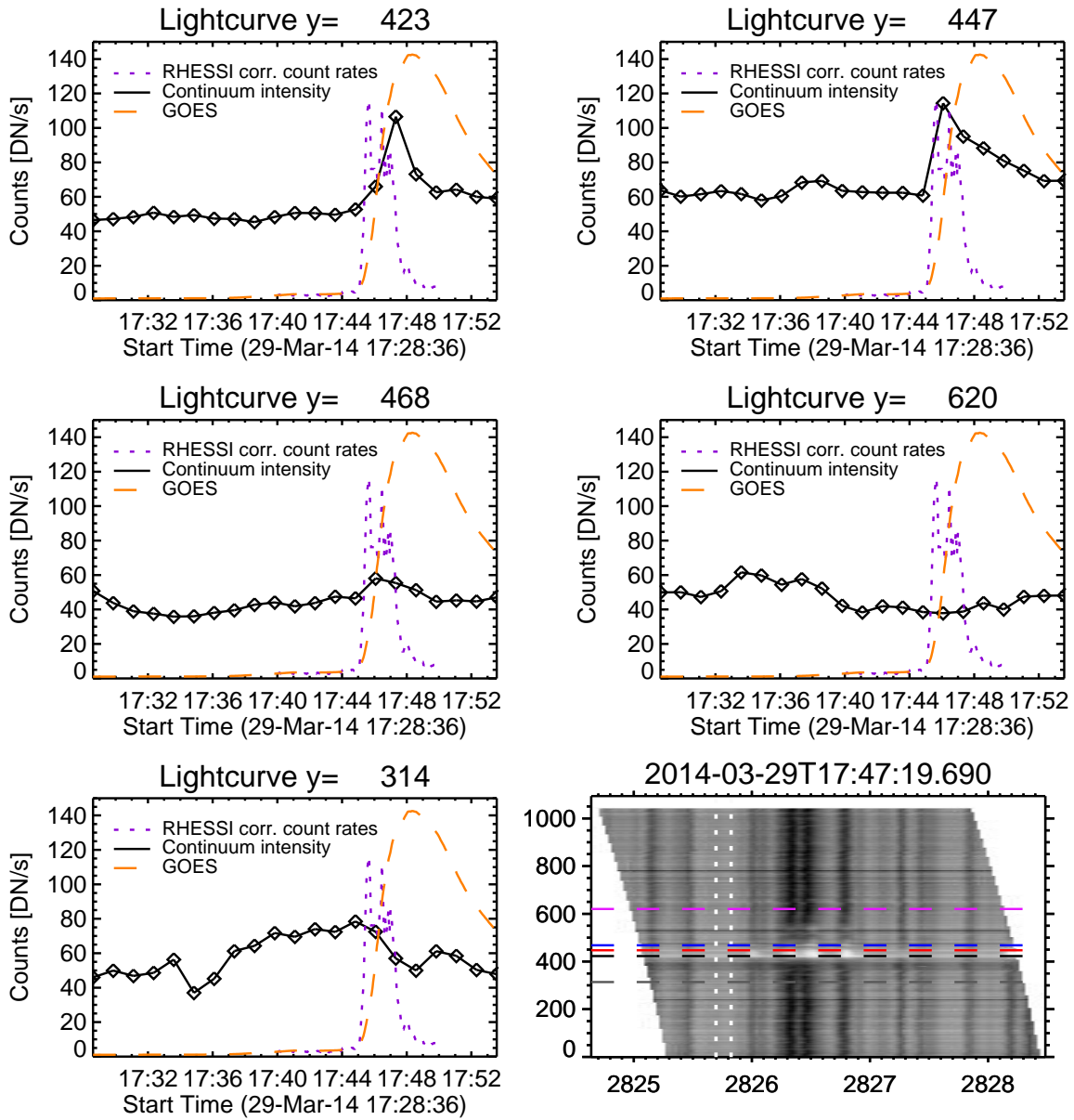}
   \caption{Lightcurves of the continuum intensity (defined as the average between the two white dotted vertical lines on the bottom right) at five positions along the slit (pixel coordinates given in title of each plot). A comparison with GOES and RHESSI lightcurves is also shown.}
         \label{fig4}
  \end{figure*}

Obviously, for a conclusive proof whether this continuum variation is due to the flare, it is necessary to investigate the temporal behavior, which is shown in Fig.~\ref{fig4}. The lightcurves in counts/second (DN/s) are plotted for each of the five positions along the slit, which are again indicated in the magnified spectrum on the bottom right. The counts were averaged over a small region of the continuum at about 2825.7 -- 2825.8 \AA\ (denoted by white vertical lines in the spectrum) and plotted for each time when the slit was at the same position (raster step 4). The lightcurves in the top row (y=423 and y=447) clearly show a strong increase in continuum counts by about a factor of two. The lightcurve from y=468 may show a very small increase too and the reference spectra (QS at y=620 and bright feature at y=314) do not show any flare-related changes. For reference, the plots also include the GOES X-ray curve (orange) and the RHESSI corrected count rates of 25-50 keV (purple), both in arbitrary units. However, these were derived by averaging over the whole active region only to show that the continuum variations are indeed flare-related while a more detailed analysis will be carried out in the future to analyze RHESSI lightcurves at each pixel.

\begin{figure*}    \centering 
   \includegraphics[width=\textwidth]{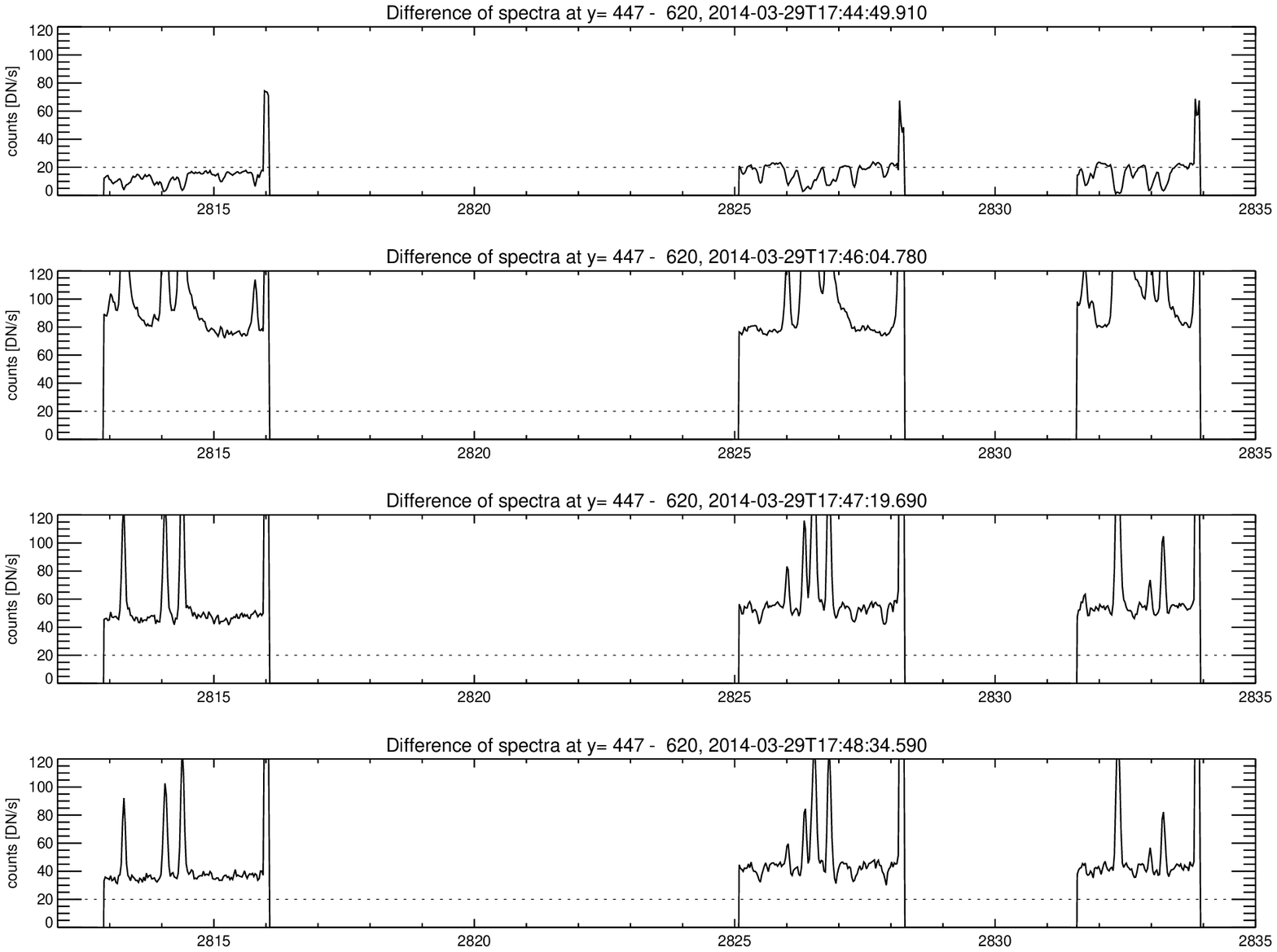}
   \caption{Difference of a spectrum in the flare ribbon (y=447 px) and the quiet Sun (y=620 px) at each timestep given in the title. A uniform increase of the continuum 
   (between lines) in the whole spectral region is clearly visible. The dashed line is drawn for reference to
 indicate the approximate pre-flare value.}
         \label{fig5}
  \end{figure*}

We can also investigate if the continuum increase shows any spectral characteristics by subtracting a QS spectrum at each timestep. From Fig.~\ref{fig4} we know that the QS did not vary significantly during the flare. For Fig.~\ref{fig5}, we therefore subtracted the QS from a spectrum in the flare ribbon (y=447) to obtain the absolute change in counts per second over the whole spectral range. Before the flare, the difference was about 20 DN/s, and it increased during the flare to over 80 DN/s. All panels of Fig.~\ref{fig5} show a uniform increase of the different continuum regions, indicating that the is no detectable spectral dependence of the continuum brightness (within this limited spectral range). The spectral lines obviously vary more. A constancy of the difference spectra indicates no brightening of the far $h$-line wing itself during the flare.

\section{Hydrogen Recombination Continuum}

Assuming the recombination model, we
can compare quantitatively the theoretical predictions with our results. For this we first
convert the observed continuum enhancement to {\em cgs units} of erg sec$^{-1}$ cm$^{-2}$ sr$^{-1}$ \AA$^{-1}$.
In Fig.~\ref{fig4} we show the lightcurves measured as the spectrum average in a narrow
'quasi-continuum' window between 2825.7 and 2825.8 \AA. The quiet-Sun level 
(y=620) is somewhat varying with time having an average level around 50 DN/s
(note that all estimates here are made for the actual position of the flare on the disk
which is $\mu$=0.83). Using recent results of HRTS-9 quiet-Sun observations,
we derived the corresponding disk intensity at $\mu=0.83$ and at our 
'quasi-continuum' window as 3.7$\times 10^5$ cgs units -
see Fig. 11 and 14a in \citet{2008ApJ...687..646M}. In their Fig. 13c, our narrow window
is also indicated as a quasi-continuum. The lightcurve at flare position y=447 peaks at
the level 115 and subtracting the pre-flare level 60 we get the pure continuum enhancement
55 DN/s which corresponds to 4.1$\times 10^5$ cgs units.

The NLTE modeling of the Balmer recombination continuum formed during flares was
performed by various authors, see a review by \citet{2007ASPC..368..417D}. Here we use the results
based on static flare models of \citet{1983ApJ...272..739R}, who constructed a grid of
models in energy balance with the electron-beam (CCTM) and conductive energy deposit
in the lower atmospheric layers. As an example, we considered their model E4 with
the electron-beam flux $F_{20}$ equal to 10$^{11}$ erg sec$^{-1}$ cm$^{-2}$ and the 
spectral index $\delta=5$. A preliminary calculation of
RHESSI energy deposition rates yields the same order of magnitude for this flare (S. Krucker, private communication). 
Using the MALI (Multilevel Accelerated Lambda Iteration) NLTE technique and including the non-thermal 
collisional rates for hydrogen, we computed the intensity of the recombination Balmer continuum
 for this specific model \citep[see also][]{heinzel2014}. At our continuum window, this intensity is equal
to 3.2$\times 10^5$ cgs units and the emitting
chromospheric layer has the optical thickness around 0.1, i.e. is optically thin as expected. 
We thus see that
this theoretical value is very much consistent with the observed enhancement
4.1$\times 10^5$ cgs units. Note that by increasing the pressure in coronal loops 
above the model chromosphere, one can achieve more than a factor of two enhancement
of the Balmer continuum within the emitting layer. Our results are also in good agreement 
with previous semi-empirical NLTE modeling of \citet{1986lasf.conf..216A} who obtained
a similar Balmer continuum enhancement for their model F2, but for very strong flares
(model F3) the enhancement is a factor of three larger. In case of optically-thin
emitting layers, the theoretical flare contrast in the Balmer continuum dramatically increases
with decreasing $\mu$ and this will help the continuum detection when the flare is observed close to the limb. 

We have also tested the absolute radiometric calibration of IRIS within the NUV window using a conversion
curve between DN/s and our cgs units, the factor
is 2.23 $\times 10^4$ which gives the QS level 1.1 $\times 10^6$ cgs units,
three times larger than the HRTS-9 value. This then implies a stronger enhancement of the
flare continuum which could be explained with other models. The IRIS calibration was performed before launch and is estimated to be accurate within a factor of two, while HRTS-9 was cross-calibrated via SUSIM/UARS, which in turn was calibrated via SOLSTICE/UARS, so both methods may have similar error bars.

\section{Discussion and Future Prospects}

Visual inspection of the IRIS spectrum in Fig.~\ref{fig2} taken just after at the flare HXR 
maximum immediately 
suggests that there is a considerable continuum enhancement at locations of strong
\ion{Mg}{2} line emission. This is evident at the three narrower spectral windows relatively far
from the the position of \ion{Mg}{2} lines.
However, there are also some other locations outside the flare region which exhibit
brighter continua, actually there are many such bright strips in this example spectrum. This is not surprising because the photospheric continuum in this
UV spectral range varies even for quiet-Sun spectra.
However, we have shown that the enhancements at flare positions
significantly exceed the other structural variations and, most critically, they vary in
accordance with the flare evolution. Moreover, at flare positions 423 and namely at 447
we have identified unambiguous impulsive increase of the continuum emission, consistent
with the HXR and GOES variations. We consider this an evidence of the continuum brightening 
at flare locations. 

This impulsive increase and a gradual decrease look qualitatively similar
to those recently found by \citet{2014ApJ...783...98K} (see their Fig. 5 and 6) for the 
15 February 2011 X2.2 flare observed in the visible continuum by Hinode/SOT. As a
potential source of the WLF emission, they consider the Paschen
continuum which is emitted by an optically-thin slab producing hydrogen recombination
radiation. The same process was considered for interpreting our observations in
the Balmer continuum. Further radiation-hydrodynamical (RHD)
simulations are required to predict the light curves of Balmer and Paschen continua,
consistently with time evolution of the electron-beam energy deposition. The latter
is usually derived from RHESSI spectra and drives the RHD simulations \citep{2010ITPS...38.2249V,2014arXiv1406.7657M}.

There might be a question whether the continuum enhancement is not an artifact
of a flare brightening in the wings of the \ion{Mg}{2} lines.
From previous observations \citep{1984SoPh...90...63L} and from our
NLTE simulations it seems to be unlikely that very far
photospheric wings of these lines (i.e.\,up to 30 \AA\ from the  \ion{Mg}{2} $h$ line center)
go into emission. Such an emission should also gradually decrease with the increasing
distance from the line center. We thus tested the excess emission in Fig.~\ref{fig5} and found
that at time of maximum continuum enhancement, it is constant over the passband
of our three windows. This suggests that we see the Balmer continuum emission added on
top of the quiet photospheric spectrum in the far wing of the MgII $h$ line.
Regarding the possible influence of the numerous metallic lines which go into emission during the flare (Fig.~\ref{fig3}), we
postpone this question to a future study where we plan to compute the complete synthetic flare spectrum in
the whole NUV window. Currently, we can only estimate a negligible contribution of the metallic line wings
to the continuum brightness. First, we saw in Section 3 that pixels 314 and 468
have the same enhanced continuum while the flare pixel shows several
metallic lines in emission. The other argument is based on synthetic spectra of the semi-empirical flare model 
F3 \citep{1986lasf.conf..216A}, where some metallic lines are in strong emission without any effect on the
surrounding continuum level. 

While we focused on the NUV spectra, IRIS also records FUV spectra from $\sim$1331--1358 \AA\ and $\sim$1380--1407 \AA. These data, which are not shown here, also show a significant brightness enhancement in the continuum, with similar timings as the NUV. But in the FUV wavelength range, there likely is a strong contribution from the optically thick continua from silicon and carbon \citep{vernazzaetal1981}, while the Balmer continuum contribution is expected to be small in the FUV during
flares. So it would be very hard to disentangle these contributions, which is why the NUV window is much better suited for the Balmer continuum detection and therefore used in this exploratory letter.

In this study we have detected a significant brightness enhancement in the IRIS NUV continuum during an X-class
flare. We exclude a possible influence of metallic line emission and show quantitatively that the observed
enhancement is quite consistent with the height-integrated emissivity of the hydrogen recombination Balmer
continuum. However, we cannot a priori exclude a possible, but according to models less probable enhancement of the photospheric continuum (quasi-continuum
in our case, represented by the far wings of the \ion{Mg}{2} lines).
Therefore, it is highly desirable to detect the Balmer continuum at least at two distant wavelengths, one
being the IRIS NUV window and the other near the Balmer limit currently detectable by some ground-based instruments.
The IRIS NUV window is very promising for future flare research, providing a unique flare diagnostics of the
Balmer continuum, as well as of many metallic lines, including  \ion{Mg}{2} $h$ and $k$.       

\acknowledgments
This work was supported by the FP-7 collaborative project No.\,606862 'F-CHROMA' (PH) and by ASI ASCR project RVO:67985815 (PH). LK was supported by a Marie-Curie Fellowship and by the NASA grant NNX13AI63G. We would like to thank Marina Battaglia for providing the RHESSI corrected count rates and the HXR contours. We also thank
Sam Krucker and the anonymous referee for valuable comments. IRIS is a NASA small explorer mission developed and operated by LMSAL with mission operations executed at NASA Ames Research center and major contributions to downlink communications funded by the Norwegian Space Center (NSC, Norway) through an ESA PRODEX contract.

\bibliographystyle{apj}
\bibliography{journals,ref}

\clearpage

\end{document}